# REFLECTION OF FEDERAL DATA PROTECTION STANDARDS ON CLOUD GOVERNANCE

Olga Dye, Justin Heo, and Ebru Celikel Cankaya

Department of Computer Science, The University of Texas at Dallas, Richardson, Texas, USA

## ABSTRACT

*As demand for more storage and processing power increases rapidly, cloud services in general are becoming more ubiquitous and popular. This, in turn, is increasing the need for developing highly sophisticated mechanisms and governance to reduce data breach risks in cloud-based infrastructures. Our research focuses on cloud governance by harmoniously combining multiple data security measures with legislative authority. We present legal aspects aimed at the prevention of data breaches, as well as the technical requirements regarding the implementation of data protection mechanisms. Specifically, we discuss primary authority and technical frameworks addressing least privilege in correlation with its application in Amazon Web Services (AWS), one of the major Cloud Service Providers (CSPs) on the market at present.*

## 1. INTRODUCTION

Cloud migration has been rapidly evolving in the recent decade. Remarkably, federal and state governments accelerated cloud adoption due to the COVID-19 pandemic. However, incidents of data breaches of the United States digital assets represent the most crucial challenge for cloud service providers (CSP) that are required to ensure resiliency from data breaches and reduce the potential for cyber threats imposed by external and internal actors.

According to the United States Cybersecurity and Infrastructure Security Agency (CISA), state-sponsored adversaries known as Advanced Persistent Threat (APT) groups, such as Russia and China, "may be seeking access to obtain future disruption options, to influence U.S. policies and actions or to delegitimize U.S. state, local, territorial, and tribal (SLTT) government entities" [28]. These ATP groups and many other adversaries continuously probe the systems for any vulnerabilities to initiate activities seeking financial gain, operational disruption, asset damage, and espionage. In addition to external cyber threats, organization employees present a variety of internal threats related to malicious activity and accidental security incidents.

It is worth noting that internal threat derives from dismantled employees who can use their authorized access to leak sensitive information. CISA defines insider threat as "the threat that an insider will use their authorized access, wittingly or unwittingly, to do harm to the department's mission, resources, personnel, facilities, information, equipment, networks, or systems." [28] As noted above, internal malicious activity can result from deliberate privilege misuse or end-user errors, which suggests that we should enhance data security practices.

According to the 2023 Verizon Data Breach Investigations Report (DBIR), privilege misuse has significantly increased in the past three years (Figure 1).

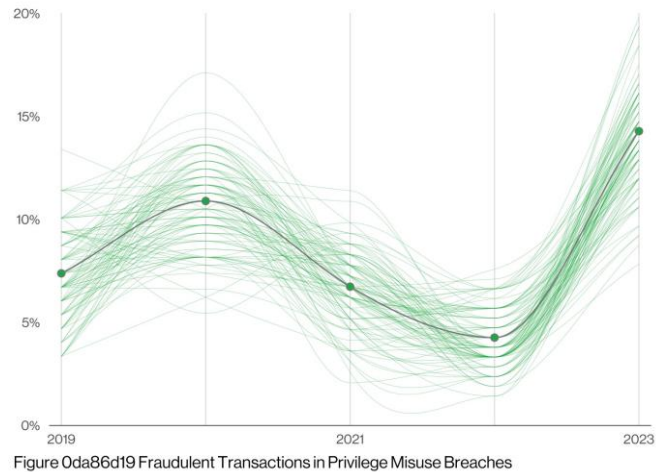

Figure 1. Privilege Misuse breaches paired with fraudulent transactions [pp. 47, 1], accessed Jun. 11, 2023. The graph shows possible values that exist within the confidence interval. The individual threads indicate a sample of all possible connections between the points within each observation's confidence interval. Here, looser threads represent a more complete confidence interval and a smaller sample size.

Additionally, the report shows that privilege misuse is one of the most common patterns causing data breaches (Figure 2). Thus, privilege misuse has led to over 406 incidents; among those - 288 incidents have led to personal and other valuable data disclosure.

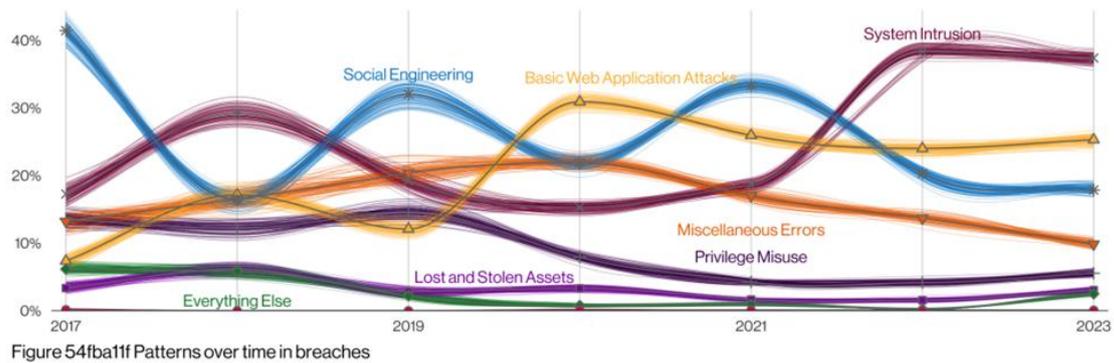

Figure 2. Breach patterns change over time. The figure demonstrates that incidents of privilege misuse are one of the most common incidents that led to confirmed data loss. Privilege misuse is predominantly driven by unapproved or malicious use of legitimate privileges [1, pp. 23], accessed Jun. 11, 2023.

Threat actors are most represented by internal actors who deliberately initiate insecure practices and are typically motivated by financial gain as demonstrated in the DBIR report and shown on Figure 3 below [1]. It is worth noting that 99% of threat actors are classified as internal [1, pp. 46].

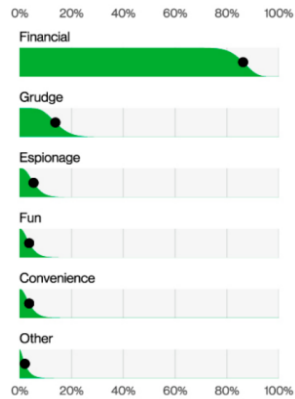

Figure 3. Financial gain is the most common motive for privilege misuse initiated by an internal actor [pp. 46, 1], accessed Jun. 11, 2023. The slant on the bar chart indicates the uncertainty of that data point to a 95% confidence level. According to the IBM report, 43% of organizations that reported data breaches were in the early stages of security practices or had yet to start applying security practices to protect their cloud environments [4]. The data breach resulted in approximately $4.53 million USD.

In addition to the monetary damages a data breach may cause, the risk of data leakage is especially significant in critical infrastructure. Since many organizations and federal agencies have transitioned their data to Cloud Service Providers (CSPs), the protection of Personally Identifiable Information (PII) and confidential data has become a long-standing concern. Thus, it is apparent that data breach incidents require CSPs to carry out more stringent assessments to reduce the potential of such instances.

Notwithstanding the security principles addressed in several frameworks and mandated by federal laws and regulations, the potential for data breaches still exists. Over-privileged policy, granting permissions on the trust basis becomes fruitful ground for internal data theft in the hands of many notorious actors. Among them is Edward Snowden, whose unlimited access to the NSA systems allowed him to obtain sensitive documents from several agencies. Perhaps, the Snowden leak urged the federal agencies and CPSs to redesign their admin privileges and revoke over-privileged permissions. Hence, a series of technical guides and regulations address the data protection methods overall and the concept of least privilege.

Incidents of data breaches of the United States' digital assets represent our motivation for this paper. CSPs are required to develop highly sophisticated mechanisms and governance that reduce data breach risks in a cloud-based infrastructure. Cloud governance by broad definition combines multiple data security measures that concatenate with legislative authority. Hence, our work lies at the intersection of two central topics: legal aspects aimed at the prevention of data breaches and technical requirements regarding the implementation of data protection mechanisms. Specifically, we will discuss primary authority and technical frameworks addressing least privilege in correlation with its application in one of the major CSPs, Amazon Web Services.

## 2. BACKGROUND AND RELATED WORK

There are many publications that focus on data protection methods in CSPs, and the corresponding data protection framework incorporated in cloud governance. Notably, many technical publications in academia are related to mining security design in CSPs, such as an implementation of the least privilege principle on various levels of data protection schemes.

Regarding governance, many publications are concentrated on data privacy and personal data protection. For example, Singh et al. noted that data protection is one of the main concerns of cloud services. Many CSP users need to know how their information is stored within the cloud, and regulations between countries fluctuate in their restrictions on data flow and privacy; some prevent specific information from leaving government borders [22].

Ali et al. highlight the role of government regulations in the adoption of cloud computing. Specifically, Ali's research is based on the survey of government regulations specific to cloud computing and presents an analysis concerning the Australian Federal Government's 'cloud-first' strategy [15]. Among many significant empirical implications, the researchers underlined that government data protection policies play an essential role in safeguarding data. In the subsequent research, Ali et al. discussed security requirements for cloud computing in the government sector [16]. Thus, the researchers identified that security controls pertaining to international standard ISO 27002 are needed to establish cloud security requirements for governments.

Ali et al. emphasized the need to ensure access controls in the proposed framework since CSPs contracting with governments are responsible for safeguarding sensitive and private data [16]. Particularly, based on the Australian regional local government context, the researchers made another interesting observation - security compliance in local government was not significantly related to cloud security [16]. The authors proposed that more action is required related to compliance issues.

Remarkably, Markopoulou et al. discussed the new European Union cybersecurity framework. This is another publication that is related to our work and addresses the subject of data protection governance in the EU [5]. In particular, the paper discusses the Network and Information Security (NIS) Directive on securing data in cloud infrastructure in the EU, the role of the European Union Agency for Cybersecurity (ENISA) in ensuring high-level information security, and the General Data Protection Regulation (GDPR) concerning the processing of personal data [5]. The paper emphasizes EU Member States' obligations with respect to national strategy. The paper provides many findings that, regardless of the choice of a legal instrument, the EU offers a cybersecurity framework while establishing compliance requirements for CSPs. With emphasis on different perspectives, the data protection framework should provide resilience to cyber threats and mechanisms, reducing the potential for a personal data breach.

The concept of least privilege has been discussed in many publications that address the security principles of CSPs. J. H. Saltzer and M. D. Schroeder were the pioneers who recognized the significance of least privilege and introduced the concept in "The Protection of Information in Computer Systems," published in July 1974 [7]. Although the novelty of the concept is no longer a debate, many studies demonstrated that maintaining the least privilege is challenging and requires various designs to address the vulnerabilities effectively.

Regarding least privilege among other data protection methods, Joshi et al. emphasize that cloud providers should provide access controls to their customers so that they can grant permissions to other authorized users as needed. This will enforce data protection and allow the implementation of security control mechanisms effectively [8].

Shimizu et al. discuss a challenge associated with retaining the least privilege on the cloud, explicitly determining a minimum set of permissions, which is an error-prone and time-consuming task for developers [21]. Hence, Shimizu discovered an approach addressing least privilege by iteratively evolving a set of permissions guided by test suites [21].

Furthermore, Soltys et al. defined the principle of least privilege as giving access to data only to those specialists who need access. According to Soltys, the approach begins by denying access to everything and allowing access as required [13].

Based on the example of Amazon Web Services, Mishra demonstrated that if a user's task is to start and stop the Amazon Elastic Compute Cloud (EC2) instances, instead of giving them full access to EC2, we should only permit them to start and stop the EC2 instance. Hence, we should always create least-privilege permissions when designing permissions [18].

Noteworthy, Gill et al. recognized the limitations of AWS's least privilege policy and identified several deficiencies in least-privilege sets. For example, a wildcard "*" in an identity-based policy may result in an over-privileged design. However, as Gill explains that even if we provide a narrower specification of the resource, the call may not succeed because in some instances the resource or the action in an identity-based policy must be the wildcard "*".

To enhance granular control over permissions in a policy, AWS recommends designing a least-privileged policy that concatenates with an attribute-based access control (ABAC). Karimi et al. proposed an automated approach regarding ABAC in another publication related to the technical aspect of data preservation mechanisms [10]. The approach was designed to help CSPs mitigate many flows in the policies and reduce the potential for data breach incidents. Specifically, the researchers proposed an unsupervised machine learning-based approach to automate ABAC policy extraction [10]. This approach works by clustering access right tuples based on the similarity of their features. The authors demonstrated that their model is robust and effective for policies with many attribute filters.

While academia continues to produce scholarly papers that address least privilege and many other data protection methods in CSPs, the legislative authority is a primary source, to which organizations must align their data protection policies. It is worth noting the key legislation the US. federal regulations and publications concerning the concept of least privilege and their effect on the CSPs' data protection mechanisms, taking AWS as an example.

## 3. ANALYSIS AND DISCUSSION

With respect to the United States' key legislation, the National Cybersecurity Strategy released on March 2, 2023, is undeniably the most critical framework, which sets forth five pillars that secure cyberspace of our nation's critical infrastructure. Specifically, in this scheme, the Biden-Harris Administration recognized a menace of malicious use and a threat of espionage in cloud-based infrastructure and discussed the commitment of the administration to work with CSPs to accelerate secure cloud migration, replace legacy systems, and modernize technology cloud security tools by adopting zero-trust architecture [24].

One of the key pieces of legislation issued prior to this strategy is the Federal Information Security Modernization Act (FISMA) of 2002. According to the act, CSPs that provide cloud services to federal agencies are obliged to follow the requirements in compliance with FISMA. Following federal compliance allows CSPs to acquire government contracts with federal agencies and improve their competitiveness in the cloud-based industry. Compliance with FISMA implies that CSPs are required to produce information security controls as defined by the NIST Special Publication 800-53 (NIST SP 500-83). Updated in 2014, FISMA identifies information security policies regarding reporting data breaches.

In addition to FISMA certification, the Federal Risk and Authorization Management Program (FedRAMP) provides authorization for cloud computing products. Recognizing the priority of a safe transition to secure CSPs and the data protection mechanisms against data breaches, FedRAMP generated a comprehensive risk assessment approach. The approach represents a combination of standards issued by the Joint Authorization Board (JAB), governing body of FedRAMP. JAB comprises the Chief Information Officers from the Department of Defense (DoD), the Department of Homeland Security (DHS), and the General Services Administration (GSA) [6]. Additionally, the FedRAMP Security Controls are issued in accordance with the NIST security standards. Moreover, FedRAMP recently passed an important milestone in January 2023, when the President signed the FedRAMP Authorization Act as part of the FY23 National Defense Authorization Act (44 U.S.C. § 3608) [6]. According to the FedRAMP authorization report, 307 cloud service providers have received authorizations within high, moderate, and low impact levels. Specifically, 132 authorizations are provided to hybrid and public CSPs, whereas the remaining are provided to cloud-based platforms designed particularly for government infrastructure [6]. Figure 5 below demonstrates major public and government CSPs with more than 20 FedRAMP authorizations.

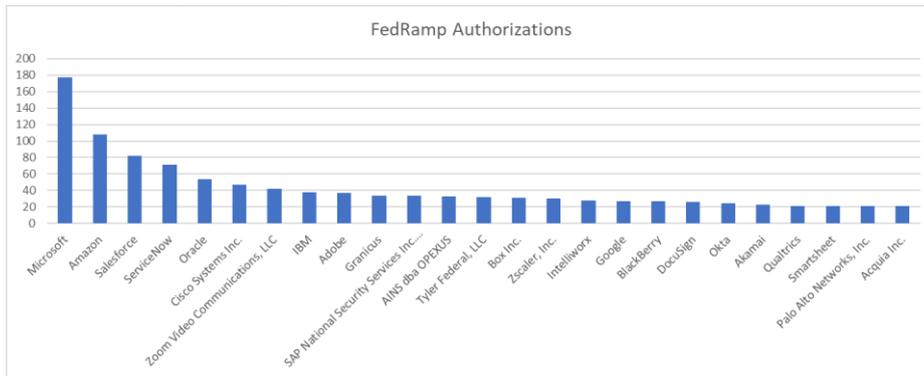

Figure 4. FedRAMP Authorizations in Cloud Service Provider [6], accessed Jun. 11, 2023.

It is imperative that this framework provides a comprehensive assessment with an emphasis on the technical guidelines regarding data protection mechanisms. We dissected the framework to comprise a practical implementation of the least privilege principle within access control requirements. Thus, in the table below we summarized essential highlights regarding least privilege standards pertaining to high and moderate baselines.

Table 1. Least privilege standards pertaining to high and moderate baselines [6], accessed Jun. 11, 2023.

| Control Name | High Baseline | Moderate Baseline |
| --- | --- | --- |
| Least Privilege | Employ the principle of least privilege, allowing only authorized accesses for users (or processes acting on behalf of users) that are necessary to accomplish assigned organizational tasks. | Employ the principle of least privilege, allowing only authorized accesses for users (or processes acting on behalf of users) that are necessary to accomplish assigned organizational tasks. |
| Least Privilege \| Authorize Access to Security Functions | Authorize access for [Assignment: organization-defined individuals or roles] to:<br>(a) [Assignment: organization-defined security functions (deployed in hardware, software, and firmware)]; and | Authorize access for [Assignment: organization-defined individuals or roles] to:<br>(a) [Assignment: organization-defined security functions (deployed in hardware, software, and firmware)]; and |
| | (b) [Assignment: organization-defined security-relevant information]. | (b) [Assignment: organization-defined security-relevant information]. |
| Least Privilege \| Non-privileged Access for Nonsecurity Functions | Require that users of system accounts (or roles) with access to [Assignment: organization-defined security functions or security-relevant information] use non-privileged accounts or roles, when accessing nonsecurity functions. | Require that users of system accounts (or roles) with access to [Assignment: organization-defined security functions or security-relevant information] use non-privileged accounts or roles, when accessing nonsecurity functions. |
| Least Privilege \| Network Access to Privileged Commands | Authorize network access to [Assignment: organization-defined privileged commands] only for [Assignment: organization-defined compelling operational needs] and document the rationale for such access in the security plan for the system. | Not Defined |
| Least Privilege \| Privileged Accounts | Restrict privileged accounts on the system to [Assignment: organization-defined personnel or roles]. | Restrict privileged accounts on the system to [Assignment: organization-defined personnel or roles]. |
| Least Privilege \| Review of User Privileges | (a) Review [Assignment: organization-defined frequency] the privileges assigned to [Assignment: organization-defined roles or classes of users] to validate the need for such privileges; and (b) Reassign or remove privileges, if necessary, to correctly reflect organizational mission and business needs. | (a) Review [Assignment: organization-defined frequency] the privileges assigned to [Assignment: organization-defined roles or classes of users] to validate the need for such privileges; and (b) Reassign or remove privileges, if necessary, to correctly reflect organizational mission and business needs. |
| Least Privilege \| Privilege Levels for Code Execution | Prevent the following software from executing at higher privilege levels than users executing the software: [Assignment: organization-defined software]. | Not Defined |
| Least Privilege \| Log Use of Privileged Functions | Log the execution of privileged functions. | Log the execution of privileged functions. |
| Least Privilege \| Prohibit Non-privileged Users from Executing Privileged Functions | Prevent non-privileged users from executing privileged functions. | Prevent non-privileged users from executing privileged functions. |

This comprehensive technical guidance of implementing least privilege is mostly similar for both high and moderate baselines but contains a few distinctions. For example, privilege levels for code execution for high baseline are defined as preventing the organization-defined software from executing at higher privilege levels than users executing the software. In addition, a high baseline restricts Network Access to Privileged Commands as follows: network access to organization-defined privileged commands can only be authorized for organization-defined compelling operational needs and such authorization should document the rationale for such access in the security plan for the system. These and many other guidelines within the access control category can help CSP customers decide whether a CSP meets the requirements for data protection.

By adhering to cloud data protection standards defined in the FedRAMP security guidelines, CSPs demonstrate their commitment to state-of-the-art data protection mechanisms, which are replicated in the form of consistent and transparent procedures.

Aligned with the US key legislation, the Executive Order on Improving the Nation's Cybersecurity provides a full scope of its authorities and resources to protect and secure its computer systems, including cloud-based service providers [15].

Following Executive Order (EO) 14028, the Office of Management and Budget (OMB) issued a memorandum that sets forth a zero-trust architecture (ZTA) mechanism and is required to be implemented by agencies by the end of 2024. ZTA has become a widely accepted approach among many CSPs. ZTA is introduced in the United States Cybersecurity and Infrastructure Security Agency's (CISA) zero trust maturity model similar to FedRAMP authorization requirements. Coherent with the five pillars: Identity, Devices, Networks, Applications, and Data, ZTA requires the CSPs to enforce granular "least privilege per-request access decisions in information systems and services in the face of a network viewed as compromised." [18, pp. 5].

In response to EO 14028, CISA, the United States Digital Service, and FedRAMP released the Cloud Security Technical Reference Architecture (TRA) [23]. The concepts underlined within the TRA bring rigorous data protection mechanisms to light. For instance, the reference defines the principle of least privilege, which encompasses sufficient privileges granted to a team member for a specific duration of access. As defined by the TRA, the principle of least-privilege "right-sizes the scope and duration of access for each person to perform the duties of their tasks and roles" [3, pp.24]. The framework recommends agencies enforce granular control and continuous monitoring privileged escalation and access to resources.

Many CSPs recommend implementing the least privilege principle whenever it is possible. For example, in its 1251-page user guide Amazon Web Services (AWS) describes the several best practices to implement the least privilege principle by: 1) providing only sufficient permissions, 2) granting additional as needed based on the role, and 3) avoiding long-term credentials whenever possible [2]. Specifically, the least privileges are achieved by the following measures: implementing AWS multi-factor authentication (MFA), creating permissions and policies in IAM, granting access across AWS accounts (IdP), creating MFA-enforced policy, and constructing policy to include tags and condition statements. In relations to MFA enforcement, some applications or external facing services can implement security through other means such as single sign-on (SSO) to prevent attackers from gaining unwanted access in their systems [29]. The condition statement in the policy compares keys in the request context with key values that we specify in our policy. For example, when we specify an AWS account, we can use the Amazon Resource Name (ARN): "Condition" : "Principal" : { "AWS":"arn:aws:iam::123456789:root" }or "Principal": {"AWS": "123456789"} [2]. We can include aws:userid and aws:SourceIdentity condition keys in a policy to revoke a federated user's active AWS sessions.

To restrict a customer-managed policy further, AWS suggests using tags [2]. These tags are attributes and are defined within ABAC. With the attribute-based approach, operations are allowed only when the principal's tag matches the resource tag, which simplifies permission management. ABAC is effective when we need to add new projects and team members.

According to NIST Cybersecurity Framework, NIST.SP.800-162, ABAC is "a logical access control methodology where authorization to perform a set of operations is determined by evaluating attributes associated with the subject, object, requested operations, and, in some cases, environment conditions against policy, rules, or relationships that describe the allowable operations for a given set of attributes" [14]. ABAC only allows actions on all resources if the resource tag matches the requester's tag. Thus, permissions within ABAC policy are limited by the number of attributes corresponding with an object or a subject.

Furthermore, TRA recognizes ABAC as "a step further by enforcing checks around the user's identity, the attributes of the resource being accessed, and the environment." Moreover, the TRA urges combining several attributes to achieve a granular policy. For example, in addition to an identity-based attribute that gives information about a user's role, an environment-based attribute can provide information about a user's device before permitting a user to perform an operation. In addition, the TRA recommends conducting security risk assessments to detect over-privileged accounts and thus, enforce the least privilege principle. [3, pp. 45]. Integration of the policy has been seen in the policies of major CPSs.

In many ways the ABAC can be used, AWS recommends implementing ABAC using the job function and project attributes as session tags when users access from a SAML IdP. To do that, we can use SAML attributes which include project assignments. Then, we need to pass these attributes as session tags to control access to the resources based on these session tags. To implement this design, AWS Skill Builder proposes a scenario in which an organization requires systems engineers to manage the EC2 instances and database engineers to manage the RDS instances. Suppose the policy requires access only to the resources related to their job function and projects. As administrators, we will need to implement ABAC using the jobfunction and project attributes as session tags, then tag project resources with the appropriate project tag. First, we create an IAM role called MyProjectResources that systems engineers and database engineers will assume when they federate into AWS to access the resources within EC2 and RDS (see Figure 6).

```
{
    "Version": "2012-10-17",
    "Statement": [
        {
            "Effect": "Allow",
            "Action": "rds:DescribeDBInstances",
            "Resource": "*" },
        {
            "Effect": "Allow",
            "Action": [ "rds:RebootDBInstance", "rds:StartDBInstance", "rds:StopDBInstance" ],
            "Resource": "*",
            "Condition": {
                "StringEquals": {
                    "aws:PrincipalTag/jobfunction": "DatabaseEngineer", "rds:db-tag/project": "${aws:PrincipalTag/project}" } } },
        {
            "Effect": "Allow",
            "Action": "ec2:DescribeInstances",
            "Resource": "*" },
        {
            "Effect": "Allow",
            "Action": [ "ec2:StartInstances", "ec2:StopInstances", "ec2:RebootInstances", "ec2:TerminateInstances" ],
            "Resource": "*",
            "Condition": {
                "StringEquals": {
                    "aws:PrincipalTag/jobfunction": "SystemsEngineer", "ec2:ResourceTag/project": "${aws:PrincipalTag/project}" } } }
    ]
}
```

Figure 6. IAM role with permissions based on attributes [2], accessed Jun. 11, 2023.

Then, create the IAM policy and attach it to the MyProjectResources role. We add a condition element based on the jobfunction and project attributes in the policy. This is to ensure that jobfunction and project tag match those assigned to the users. Next, to ensure that the systems engineers and database engineers can assume this role when they federate into AWS from the IdP, modify the role's trust policy to trust the SAML IdP. After that, we need to add a new action sts:TagSession in the policy statement which includes session tags when engineers federate in. The policy must also include a condition that requires the jobfunction and project attributes to be included as session tags when engineers assume this role. (see Figure 7).

```
{
    "Version": "2012-10-17",
    "Statement": [
        {
            "Effect": "Allow",
            "Principal": { "Federated": "arn:aws:iam::999999999999:saml-provider/ExampleCorpProvider" },
            "Action": [ "sts:AssumeRoleWithSAML", "sts:TagSession" ],
            "Condition": {
                "StringEquals": {
                    "SAML:aud": "https://signin.aws.amazon.com/saml" },
                "StringLike": {
                    "aws:RequestTag/project": "*",
                    "aws:RequestTag/jobfunction": [ "SystemsEngineer", "DatabaseEngineer" ] }
            }
        }
    ]
}
```

Figure 7. Adding the new action in the policy statement, accessed Jun. 11, 2023.

Finally, an administrator configures the SAML IdP to include the job function and project attributes as session tags in the SAML assertion when engineers federate in using this role. That way, when systems and database engineers federate in the AWS resources using, for example, the MyProjectResources role, they access only those project resources that match the project and jobfunction attributes passed in their federated session. Hence, session tags allow to specify granular permissions according to a user's attributes.

As the policies are periodically updated, the security framework includes new features. Similarly, CSPs frequently redesign their security features and improve their data protection solutions to outpace constantly evolving identity-based cyberattacks. For example, in March 2023, AWS added new condition statements to strengthen least privilege. The following keys allow us to restrict the policy and implement more granular control: aws:EC2InstanceSourceVPC and aws:EC2InstanceSourcePrivateIPv4. The first key aws:EC2InstanceSourceVPC requires that the request to a service must originate from the same instance that the credentials of the instance are issued to. Comparatively, the second key requires that the request must pass through a VPC endpoint. AWS recommends adding these condition keys to the policy below to restrict access to S3 bucket. Thus, the policy will deny access to EC2 resources if the request does not originate from the same VPC and IP of the EC2 instance (Figure 8).

```json
{
  "Version": "2012-10-17",
  "Statement": [
    {
      "Effect": "Deny",
      "Action": "s3:*",
      "Principal": {
        "AWS": "*"
      },
      "Resource": [
        "arn:aws:s3:::<DOC-EXAMPLE-BUCKET>/*",
        "arn:aws:s3::: <DOC-EXAMPLE-BUCKET>"
      ],
      "Condition": {
        "StringNotEquals": {
          "aws:ec2InstanceSourceVPC": "${aws:SourceVpc}"
        },
        "Null": {
          "aws:ec2InstanceSourceVPC": "false"
        },
        "BoolIfExists": {
          "aws:ViaAWSService": "false"
        }
      }
    },
    {
      "Effect": "Deny",
      "Action": "*",
      "Principal": {
        "AWS": "*"
      },
      "Resource": [
        "arn:aws:s3::: <DOC-EXAMPLE-BUCKET> /*",
        "arn:aws:s3::: <DOC-EXAMPLE-BUCKET>"
      ],
      "Condition": {
        "StringNotEquals": {
          "aws:ec2InstanceSourcePrivateIPv4": "${aws:VpcSourceIp}"
        },
        "Null": {
          "aws:ec2InstanceSourceVPC": "false"
        },
        "BoolIfExists": {
          "aws:ViaAWSService": "false"
        }
      }
    }
  ]
}
```

Figure 8. Customer-managed policy with new condition keys [11], accessed Jun. 17, 2023.

The abovementioned policy illustrates how CSPs continue analyzing security features and strive to deliver state-of-the-art data protection mechanisms. AWS and other prominent CSPs aim to redesign the security framework to reflect organizational changes and evolving needs of personnel and service accounts. These and many other security practices are consistent with the FedRAMP high baseline requirements for implementing least privilege. This observation allows us to premise that future security updates in cloud services will align with the National Cybersecurity Strategy.

Thus, we can speculate that FedRAMP authorization requirements set forth many updates in data protection mechanisms. Because CSPs adhere to FedRAMP, security practices become more transparent and publicly available, allowing more CSPs to adopt similar practices.

The CSPs rapidly respond to cyber threats and improve resiliency and overall data protection as identified by the recently released security framework, which essentially has one specific objective: to protect data of our nation's critical infrastructure. This tendency reflects the NIST core principles to identify, protect, detect, respond, and recover while preserving the integrity, confidentiality, and availability of data stored in cloud services. This requirement is reflected in ZTA, and other principles allowing to secure data and prevent unauthorized access.

The zero-trust architecture outlined in the Cloud Security Technical Reference Architecture, the CISA's zero trust maturity model, and other laws and regulations set the direction of data security policy and reinforced the National Cybersecurity Strategy. Furthermore, the CSPs' updated security features demonstrate moving forward with implementation of the Biden-Harris administration's objectives in cyberspace. Hence, we are likely to see the reflection of the strategy in the CSPs' future data security framework.

## 4. CONCLUSION AND FUTURE WORK

In this work, we present cloud security efforts by focusing on cloud governance, which harmoniously combines multiple data security measures with legislative authority. We present legal aspects aimed at the prevention of data breaches, as well as the technical requirements regarding the implementation of data protection mechanisms. Specifically, we discuss primary authority and technical frameworks addressing least privilege in correlation with its application in AWS. Our analysis demonstrates that the principle of least privilege finds an extensive interpretation and implementation in AWS for supporting enhanced security, including ABAC, zero-trust architecture, and condition keys.

There is a tremendous scope for future work. In this paper we observed the process of implementing least privilege in AWS policy on the example of session tagging and condition keys. Our analysis demonstrates an integration of federal data protection standards in CSPs policies that follow the scope of technical publications aligned with the National Cybersecurity Strategy and the United States key legislation designed to protect the data of our critical infrastructure. Furthermore, collaboration between government, academia, and industry must play a central role in the enhancement of cloud security mechanisms and design a defense-in-depth framework protecting data and preventing unauthorized access.

Undoubtedly, with changing governments, and ever-evolving cyber technologies including in cloud infrastructure, the fundamental security principles will also change, and will need to be updated. So, our future work is obvious, with well-defined determinant parameters, though their values remain unknown for the time being.

### ACKNOWLEDGMENTS

We would like to highlight the support from our colleague, Diana Cates during data collection efforts for our paper.

## AUTHORS

**Olga Dye** is a soon-to-be graduate with a master's degree in Cybersecurity, Technology and Policy at the University of Texas in Dallas. She previously graduated from the faculty of Chinese Philology in the Institute of Asian and African Studies at Lomonosov Moscow State University, Russia. Her interests encompass cybersecurity, public policy, and linguistics. 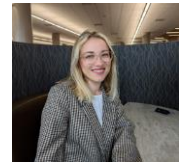

**Justin Heo** is a graduate student in the master's program of Cybersecurity, Technology and Policy at The University of Texas at Dallas. He has transitioned into the cybersecurity field from his initial focus in biological sciences at The University of Texas at Austin. His interests lie in security compliance, vulnerability management, and application security. 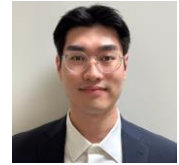

**Ebru Cankaya** is a professor of Instruction at The University of Texas at Dallas Department of Computer Science. She has received her PhD degree from Ege University in Turkey, and MS in IT and Management from The University of Texas at Dallas. Dr. Cankaya has been teaching data and cybersecurity courses at undergraduate and graduate levels, and is collaborating with students in research in these areas. 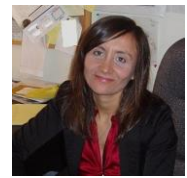